body of paper

\documentstyle[preprint,aps]{revtex}

\begin{document}

\draft

%
%

\title{Cosmic Colored Black Holes}
\author{Takashi Torii\thanks{electronic
mail:64L514@cfi.waseda.ac.jp},
Kei-ichi Maeda\thanks{electronic mail:maeda@cfi.waseda.ac.jp},
 Takashi Tachizawa\thanks{electronic mail:63L507@cfi.waseda.ac.jp}}
\address{Department of Physics, Waseda University,
Shinjuku-ku, Tokyo 169, Japan}
\date{\today}
\maketitle
\begin{abstract}
We present spherically symmetric static solutions (a particle-like
solution and a black hole solution) in the Einstein-Yang-Mills
system
with a cosmological constant.
Although their gravitational structures
are locally similar to those of the Bartnik-McKinnon particles or
the colored
black holes, the asymptotic behavior
becomes quite different because of the existence of a cosmological
horizon.
We also discuss their stability by means of a catastrophe
theory as well as a linear perturbation analysis
and find the number of unstable modes.
\end{abstract}

%
%

Self-gravitating structures with non-Abelian fields
have been intensively studied from  various points of view
since Bartnik and McKinnon
(BM) particle\cite{Bar} and colored black hole solutions\cite{Vol}
were found by the numerical
method.
Since such new types of solutions with
non-Abelian hair show a variety of features
and have many interesting properties,
studying them may reveal new important
aspects in general relativity such as the
no hair conjecture or a  stability analysis via a catastrophe
theory\cite{Tor2,Tachi}.
Those objects  may be very small even if they exist and
they would be important only
in the early history of the universe\cite{Moss,Gib}.
In the early universe, however, we usually expect
a vacuum energy, which is equivalent to a cosmological constant.
{}From an observational cosmological point of view, some
astrophysicists
 have pointed out that a small cosmological constant may explain
the observed number counts of galaxies\cite{Fuk}.
We therefore have wondered why a self-gravitating non-Abelian
structure with
a cosmological
constant has not been studied yet.
Such non-Abelian structures might have been formed in the early
universe and might have played an important role in cosmology.
Hence, it may be worthwhile studying such objects.

In this short note we
present both particle-like and black-hole  non-Abelian solutions
 with a cosmological constant.
Here we will consider only a localized object, such as the BM solution
or the colored black hole, in the universe with cosmological constant.
Then we assume the cosmological horizon, inside of which a localized
object exist. Then the spacetime approaches to the Schwarzschild-de
Sitter solution or the Reissner-Nordstr\"om-de Sitter (RNdS) solution
asymptotically near the cosmological horizon. Hence a new solution is a
direct generalization of a RNdS solution from the U(1) gauge field to
the SU(2) gauge field\cite{DH}.
We study structures and
thermodynamical properties of new solutions and
also discuss
their stability by means of a catastrophe theoretical method
as well as a linear perturbation analysis.

%
%

We start with the following Lagrangian,
\begin{equation}
  S = \int d^4x \sqrt{-g} \left[\frac1{16\pi G} (R
      -{\Lambda})
      -\frac1{16\pi g^2} \mbox{Tr} F^2
        \right],
       \label{10}
\end{equation}
where the SU(2) Yang-Mills(YM) field strength $F$ is expressed by its
potential $A$
as $F=dA+A\wedge A$, $g$ is a coupling
constant of the YM field.

Since we are interested in the case
where the YM field is localized and the spacetime
approaches asymptotically the de Sitter solution because of a
cosmological constant $\Lambda$,
the metric of the spherically symmetric static spacetime is
written as
\begin{equation}
  ds^2 = -\left(1-\frac{2Gm(r)}r-\frac{\Lambda}3 r^2\right)
         \mbox{e}^{-2\delta(r)}dt^2
         +\left(1-\frac{2Gm(r)}r-\frac{\Lambda}3 r^2\right)^{-1}dr^2
         +r^2 (d\theta^2+ \sin^2\theta d\phi^2).
 \label{20}
\end{equation}

As for a SU(2) YM potential, we adopt the following form:
\begin{equation}
  A = w(r)\tau_1 d\theta +\left\{ w(r)\tau_2+\cot \theta \tau_3
  \right\} \sin \theta d\phi.   \label{70}
\end{equation}
which is the same as that of the colored black hole and
is obtained from the most generic spherically symmetric one
with the ansatz of being static and having no ``electric" charge.

The field equations derived from (\ref{10}) are
\begin{eqnarray}
& &
   \tilde{m}' = \left(1-\frac{2\tilde{m}}{\tilde{r}}
      -\frac13 \tilde{r}^2 \right)
     \frac{w^{\prime 2}}{\alpha^2}
     +\frac{(1-w^2)^2}{2\alpha^2 \tilde{r}^2},
                \label{90}  \\
& &
\delta ' = -\frac{2w^{\prime 2}}{\alpha^2 \tilde{r}},
                \label{100}  \\
& &
\left[ \left(1-\frac{2\tilde{m}}{\tilde{r}}-\frac13 \tilde{r}^2
\right)
   {\rm e}^{-\delta}
w'\right]'
 - \frac{w(1-w^2)}{\tilde{r}^2} = 0,
                \label{110}
\end{eqnarray}
where we have used dimensionless variables normalized by
$\Lambda$ and $G$
as $\tilde{r}\equiv \sqrt{\Lambda} r$,
$\tilde{m}\equiv \sqrt{\Lambda} Gm$. A prime denotes a
derivative with
respect to $\tilde{r}$. $\alpha \equiv g/\sqrt{G\Lambda}$ is a
normalized coupling
constant, which is only one parameter appearing in those basic
equations.

%
%

We consider two classes of solutions: One is a particle-like
solution
(first class) such as the BM solution and the other is a black hole
solution (second class) such as the colored black hole.
Before showing our results, we
have to discuss boundary conditions of the field functions.
Because of the asymptotic structure of our spacetime, we expect
a cosmological horizon. Hence we need two boundary conditions
both at the origin ($r=0$) or at the black hole horizon $(r_h)$
and at the cosmological horizon $(r_c)$.
As for the metric functions, $m(r)$ and $\delta (r)$,
those must be finite in the whole region. We set $\delta =0$ at the
cosmological horizon. For the YM field, a regularity
at the origin requires $w=\pm 1$ and $w'=0$.
We can choose $w=1$ without loss of generality.
At the black hole horizon, we use the same boundary
condition as that of the colored black hole, i.e.,
$w'(r_h)$ is described by $w(r_h)$ from the condition for a
regular horizon and $w(r_h)$ becomes a
shooting parameter.
At the cosmological horizon, although a value
of $w$ seems to be free in the present coordinate system,
it must be analyzed more carefully.
It is useful to use a new coordinate $\chi$ defined by $r=R\sin \chi$,
($\chi\in [0, \pi/2]$), where $R$ is a radius of the cosmological horizon
locating at $\chi = \pi /2$. The boundary condition at $\chi = \pi /2$ is
expressed by $dw/d\chi =0$ because the spacetime approaches to the
RNdS spacetime asymptotically near the cosmological horizon.
However, because the YM equation in terms of $\chi$ becomes singular
at
the cosmological horizon ($\chi = \pi /2$), we use the previous $r$-
coordinate and transform the $r$-coordinate to the tortoise coordinate
($r^*$), defined by $dr/dr^* = 1-2Gm/r-\Lambda r^2/3$, only near the
cosmological horizon. It shifts the cosmological horizon away to
infinity ($r^{*} \to \infty$) and makes it easy to impose the regularity
at the cosmological horizon. Checking the boundary condition
at the cosmological horizon in terms of the $\chi$ coordinate as well,
we found new solutions numerically.
We shall discuss a particle-like solution and a black hole
separately.

\noindent 1. Particle-like solution

Under the above boundary conditions, we have a trivial
analytic
solutions: $w \equiv 1$, $m \equiv 0$, $\delta \equiv 0$, which is the
de Sitter solution.

Non-trivial particle-like solutions we have found are some kind of
extension of the BM solution.
We show the profiles of the field functions in Fig. 1
In the region where the YM
field is located, the new solution
is very similar to the BM solution. We find a family of discrete
solutions each of which is characterized by the node number $i$ of
the YM
potential.
The asymptotic behaviors, however, become quite different from
those of the BM solution because of the existence
of $\Lambda$. The YM field of the BM solution damps faster
than $\sim r^{-2}$ and it has no global charge relating to the gauge
field.
This is
the reason why we classified the BM solution into
a globally neutral type in the previous paper\cite{Tor2}.
In the case of the new solutions, however, the YM field does not
vanish and continues to
exist over the cosmological horizon. This produces an effective
charge
at $r=r_c$
defined by $Q_{eff} = \int_{r_{c}} \sqrt{{\rm tr} F^2} r^2 \sin
\theta
d\theta d\phi$,
hence, we expected that the spacetime
approaches to the RNdS spacetime asymptotically.
We have also found that the effective charge
gets large when the normalized coupling constant $\alpha$ becomes
small, i.e.,
when $g$ decreases or $\Lambda$ increases.

A new family of solutions has a critical coupling constant
$\alpha_{cr}\sim 1.75$,
below
which no solution exists except for trivial ones.
That is to say a non-trivial solution disappears when $\Lambda$
gets large and/or $g$ gets small.
We can easily understand this as follows.
For a general relativistic fluid with nonzero vacuum energy density
$\rho_{vac}$
(or a cosmological constant $\Lambda$),
if $\rho_f /\rho_{vac} \leq 2$,
the perfect fluid cannot be localized as an isolated star-like
object, where $\rho_f$
is a fluid density\cite{His}.
For our new solution, we shall introduce the mean energy density of
the YM field,
$\bar{\rho}_{YM}
\equiv M/(\frac43 \pi r_0^3)$. $r_0$ is the effective radius at
which $\rho_{YM}$
of the YM field drops by half. Comparing $\bar{\rho}_{YM}$ with
$\rho_{vac}=\Lambda /8\pi G$,
we find that a new solution does not exist for the case of
$\bar{\rho}_{YM}/\rho_{vac} \mbox{\raisebox{-1.ex}{$\stackrel
{\textstyle<}
{\textstyle \sim}$}} 5$.
This result is consistent with the perfect fluid case and we
expect that this property may be universal for any matter with a
cosmological
constant. The physical reason why there is the critical value
$\alpha_{cr}$
may be explained as follows: The size of the self-gravitating non-
trivial structure
is $r_0 \sim \sqrt{G}/g$, while the radius of cosmological
horizon is $r_c \sim (3\Lambda)^{1/2}$. Then, if $r_0 > r_c$, i.e.,
$\alpha \mbox{\raisebox{-1.ex}{$\stackrel
{\textstyle<}
{\textstyle \sim}$}} O(1)$, no particle-like solution can exist in
the
de Sitter background spacetime.

%
%

\noindent 2. Black hole solution

Now we turn to the black hole solutions.
It is easy to check that the Schwarzschild-de Sitter solution ($w
\equiv \pm 1$)
and the RNdS solution ($w \equiv 0$) are
trivial solutions.

As for a non-trivial black hole solution (we call it the cosmic
colored
black hole),
we plot the mass-horizon radius relation
in Fig. 2.
Note that the mass of a black hole $M$ is defined by
\begin{equation}
    M=m(r_{c}) +\frac{Q_{eff}^2}{2r_{c}}.
\label{130}
\end{equation}
The reason is as follows: A new
solution has an effective charge at the cosmological $r_c$
horizon and it approaches the RNdS
spacetime asymptotically.
Then the mass function $m$ includes a contribution of a gauge field.
It is plausible to subtract it in the definition of the mass of a
black hole just as in the RNdS solution.
Furthermore, this definition of $M$ provides the conserved AD mass
for
the RNdS black hole, although it is not
certain whether this is true for the present cosmic colored black
holes.

For the RNdS solution,
there are in general three horizons, an inner horizon,
a black hole horizon and a cosmological horizon
(the dotted lines in Fig. 2).
In Fig. 2,
the solid lines denote the black hole and the cosmological horizons
($r_h < r_c$) of the cosmic colored black holes.
In the limit of $r_h \to 0$,
these branches end up with particle-like
solutions.
The behaviors of solutions
depend on the coupling constant $\alpha$. For a large coupling
constant
$\alpha > 2.3$, the black-hole branch reaches that of
RNdS solutions, and makes a bifurcation point. This behavior
is similar to a family of monopole black holes in the
Einstein-Yang-Mills-Higgs system
\cite{Tachi}, where a family of monopole black holes merges
to the Reissner-Nordstr\"om black hole branch.
The monopole black hole has interesting properties
depending on their self-coupling constant $\lambda$ such that there
are two
types of non-trivial solutions for small $\lambda$,
one of which is more stable than the
other.
One may expect that the cosmic colored black hole has similar
properties, but
this turns out to be forbidden by catastrophe theory.
This is so because a family of monopole
black holes constructs a swallow tail catastrophe, which needs at
least
three independent
parameters, while the present system has only two parameters,
i.e., $\alpha$ and a radius of the black hole horizon.

For a small coupling constant $\alpha <2.3$,
the cosmic colored black hole solution is not
bound up with the RNdS branch but disappears on the way.
We can understand the reason for this by looking at the
branch of the cosmological horizon.
The end point of the branches merges to the extreme line
of the RNdS black holes, where the black hole horizon and
cosmological horizon coincide with each other.
When the mass of a cosmic colored black hole gets large, the radius
of its
cosmological horizon would become smaller than that of extreme RNdS
black holes. However, although a cosmic colored black hole is
different from the RNdS
black hole, the asymptotic behavior should be the same.
Hence it is likely for cosmic colored black holes to disappear at
the extreme
point of the RNdS black hole.
When the coupling constant gets smaller even further
($\alpha < \alpha_{cr} \sim 1.75$),
the cosmic colored black hole
does not exist as the particle-like solution.

%
%

To discuss the thermodynamical properties of the cosmic colored
black hole,
we plot temperatures at both black hole and cosmological horizons
of the cosmic colored black holes in Fig. 3. The temperature at the
cosmological horizon has a similar mass-dependence to that of the
RNdS
black hole qualitatively, while there is a big difference for that
at the
black hole
horizon. In the RNdS black hole case, when the mass of a black hole
gets small,
a sign of the heat capacity at the black hole horizon
changes from negative to positive, and the temperatures at both
horizons
coincide  at a point $E$ in Fig. 3. In the limit of the extreme
black hole,
the temperature vanishes. On the other hand, for the cosmic colored
black hole
when the mass of the black hole
gets small, its temperature diverges though
there is some range where the sign of its heat capacity becomes
positive.
This behavior is similar to that of the colored black hole.
This may be understood by the same mechanism explained in Ref.
\cite{Tor}.

Those thermodynamical properties may allow us to discuss the
evolution
of the cosmic colored black holes.
If there is initially a RNdS black hole whose mass is large enough,
its mass
gradually decreases via the Hawking radiation because the
temperature
at its black hole horizon is higher than that at the cosmological
horizon.
At the point $E$ in Fig. 3 where the two temperatures
eventually become equal, however,
the energy fluxes from both horizons balance
and the black hole does not evaporate further.
On the other hand, this scenario cannot be applied to the SU(2) YM
system,
because of the existence of the cosmic colored black holes.
The cosmic colored black hole has larger entropy
than the RNdS black hole with the same mass.
Hence, the RNdS black hole shifts to the cosmic colored black hole
at the bifurcation point. After this,
since there is no intersection of two temperature curves unlike in
the RNdS case,
the evaporation does not stop, but rather it will
be accelerated,
and finally a particle-like solution will remain.

%
%

Are the cosmic colored black holes stable
or not? Here we use two methods in order to answer this question.
One is the catastrophe theoretical
method\cite{Thom}, which is useful in discussing a relative
stability
among several families of
solutions and is widely applied in various research fields including
astrophysics.
The other is the usual linear perturbation method, with which we can
find
unstable modes explicitly and show
the number of such
modes.

First we show the former method. We choose the mass
$M$ and the entropy $S$ of the black hole as a control parameter and
a
potential function in the catastrophe theory,
respectively. The entropy $S$ is related to a radius of the black
hole horizon
as $S=\pi r_h^{~2}$, hence $r_h$ is qualitatively equivalent to the
entropy
$S$.
We show the $M-r_h$ relation for
$\alpha =0.1$ in Fig. 4. $B_i$ ($i=1,2$)
describe the bifurcation points consisting of branches of
cosmic colored black holes with $i$ modes
and of the RNdS black hole, and $M_i$ ($i=1,2$) are masses of
black holes at these bifurcation points $B_i$. The structure
at each bifurcation point
in a plane spanned by the control parameter and the potential
function is
classified into a cusp catastrophe.
Since the cosmic colored black hole has larger entropy
than
the RNdS black hole,
the cosmic colored black hole is more stable than the RNdS black
hole with the
same mass by
means of catastrophe theory.
Provided that
the RNdS black hole with $M>M_1$ has $n$ unstable
modes, it will find another unstable mode at the bifurcation point
$B_1$ and
then has
($n+1$) unstable modes in the range of $M_2<M<M_1$. On the other
hand, the
cosmic
colored black hole with one node has $n$ unstable modes. Similarly,
at the point $B_2$, the RNdS black hole will get
another unstable mode while the cosmic colored black hole with two
nodes has
($n+1$) unstable modes, and so forth.
The particle-like solution  has the same number of unstable modes
in its branch.
Note that if $n=0$, i.e., the RNdS
black hole with $M>M_1$ is stable, then the cosmic colored black
hole
with one node is also stable. Hence we only have to investigate the
stability
of the RNdS black hole. However it is impossible to study it by the
catastrophe
theory, and therefore we apply the linear perturbation method.
Before showing our results it should be stressed that the RNdS
black hole with a U(1) gauge field is stable!

Here we consider only radial perturbations. Writing down the
perturbation
equations and drawing their potential form, we can see that a
cosmological constant has a tendency to stabilize the unperturbed
solution.
Analyzing the stability in detail numerically, however, we find that
the RNdS solution with $M>M_1$ has one unstable mode. We have also
confirmed that
the number of
unstable modes of the RNdS black hole
increases one by one at the bifurcation points ($B_i$).
Hence we conclude that the cosmic colored black hole with $i$ nodes
has $i$ unstable
modes. This result does not depend on the coupling constant
$\alpha$.

%
%

In this paper we have investigated
a particle-like solution with a cosmological constant and
the cosmic colored black hole,
which are the first self-gravitating non-Abelian structures
with a cosmological constant.
The gravitational structure is definitely changed by the
cosmological
constant, in particular an effective charge appears at the
cosmological horizon.
Although the new solution is not stable,
we may expect some important effects in the astrophysical process
caused
by the BM solutions and/or the colored black holes\cite{Gib}.


This work was supported partially by the Grant-in-Aid for Scientific
Research Fund of the Ministry of Education, Science and Culture
(No. 06302021 and No. 06640412),
by the Grant-in-Aid for JSPS Fellows (No. 053769),
and by the Waseda University Grant
for Special Research Projects.


\newpage
\begin{flushleft}
{ Figure Captions}
\end{flushleft}

\vskip 0.1cm
   \noindent
\parbox[t]{2cm}{ FIG. 1:\\~}\ \
\parbox[t]{14cm}
{The YM potential $w$ and metric functions $m$, $\delta$ of particle-
like solutions for $\alpha = g/\sqrt{G\lambda} =2.5$ (solid lines) and
$= 4.0$ (dashed-lines) in terms of $\chi$. $\chi = \pi /2$ corresponds
to the cosmological horizon. These solutions have one node in the half-
sphere ($0\leq \chi \leq \pi/2$). The behaviors of the functions are
similar to the BM solution near the origin. From this figure, we find
that derivatives of each function with respect to $\chi$ vanish at the
cosmological horizon and the spacetime has a reflection symmetry.
}\\[1em]
\noindent
\parbox[t]{2cm}{ FIG. 2:\\~}\ \
\parbox[t]{14cm}
{The mass-horizon radius diagrams of the cosmic colored black holes
for  $\alpha\equiv g/\sqrt{G\Lambda} =10.0$($a$(the black hole
horizon(BH)),
$a'$(the cosmological horizon(CH))), =4.0($b$(BH), $b'$(CH)),
=2.2($c$(BH), $c'$(CH)), and  =1.9($d$(BH), $d'$(CH)).
We also plot those of the RNdS black holes for their
charges $\sqrt{G\Lambda}Q$ = 0.1 and 0.25 (the dotted lines) and the
extreme points of all RNdS black holes (the dot-dashed line).
We find that the cosmic colored black holes for $a, a'$ and $b,b'$
coincide with
the RNdS black holes with the same charge at the cosmological
horizons,
while the cosmological horizons for $c'$ and $d'$  finish on the
extreme line.
}\\[1em]
\noindent
\parbox[t]{2cm}{ FIG. 3:\\~}\ \
\parbox[t]{14cm}
{The mass-temperature diagram of the cosmic
colored black holes for $\alpha = 10.0$ (solid lines),
=4.0  (dotted lines), and   =2.0
(dot-dashed lines). We also plot those for the RNdS black holes with
the same charges.
For the large coupling constant $\alpha$ (e.g. 10.0),
the specific heat will change its sign two times, but
there are no changes for small
$\alpha$ (e.g. 2.0).
The temperatures of both black hole and cosmological
horizons of RNdS black holes intersect
at the point $E$, but the cosmic colored black holes do not have
such a point. This will change the fates of two types of black
holes.
}\\[1em]
\noindent
\parbox[t]{2cm}{ FIG. 4:\\~}\ \
\parbox[t]{14cm}
{The mass-horizon radius diagram of the cosmic
colored black holes with node number $i=1, 2$ and the RNdS
black holes for $\alpha=10.0$.
There are two bifurcation points $B_i$ $(i=1, 2)$, which correspond
to the cusp catastrophes. The solid and the dotted
lines have one and two unstable modes, respectively.
}\\[1em]
 \noindent


\begin{thebibliography}{99}
  \bibitem{Bar} R. Bartnik and J. McKinnon, Phys. Rev. Lett. {\bf
61}, 141
  \bibitem{Vol} M.S. Volkov and D.V. Galt'sov, Pis'ma Zh. Eksp.
Teor. Fiz. {\bf
                50} 312 (1989); Sov. J. Nucl. Phys. {\bf 51}, 747
(1990);
                P. Bizon, Phys. Rev. Lett. {\bf 64} 2844, (1990);
                H.P. K\"unzle and A.K. Masoud-ul-Alam, J. Math.
Phys. {\bf 31},
                928  (1990).
\bibitem{Tor2} K. Maeda, T.  Tachizawa, T. Torii and T. Maki, Phys.
Rev. Lett.
                {\bf 72} 450, (1994)
                See also T. Torii, K. Maeda and T. Tachizawa,
                Phys. Rev. D {\bf 51} 1510, (1995), and references therein.
 \bibitem{Tachi} T. Tachizawa, K. Maeda, and T. Torii, WU-AP/40/94
(gr-qc9410016)
              (1994, to be published in Phys. Rev. D).
  \bibitem{Moss}  I. G. Moss and A. Wray, Phys. Rev. D {\bf 46}
R1215 (1992).
  \bibitem{Gib} W. G. Gibbons and A. R. Steif, Phys. Lett. B {\bf
314} 13 (1993);
                Phys. Lett. B {\bf 320} 245 (1993);
                M. S. Volkov, Phys. Lett. B {\bf 328} 89 (1994);
                {\bf 334} 40 (1994).
  \bibitem{Fuk} M. Fukugita, F. Takahara, K. Yamashita and Y.
Yoshii,
                Astrophys. J. {\bf 361} part2 L1 (1990).
  \bibitem{DH}  It should be noted that there are also another type of
solutions in the same system for a specific value of a cosmological
constant; S Ding and A Hosoya, TIT/HEP-242/COSMO-39 (1993); P\'al
G\'eza Moln\'ar, ZU-TH7/95 (gr-qc9503036) (1995).
Those solutions are analog to the static Einstein's universe and its
excitation, and then they are globally static in contrast to our new
solution which is not globally static because of the existence of a
cosmological horizon.
  \bibitem{His}  W. A. Hiscock, J. Math. Phys. {\bf 29}, 443 (1988)
   \bibitem{Tor} T. Torii and K. Maeda, Phys. Rev. D {\bf 48}, 1643
(1993).
  \bibitem{Thom} R. Thom, {\it Structural Stability and
Morphogenesis}, Benjamin (1975).
\end{thebibliography}
\end{document}